%% file: main.tex
\documentclass{article}

\usepackage[preprint]{neurips_2021}

\usepackage[utf8]{inputenc} 
\usepackage[T1]{fontenc}    
\usepackage{hyperref}       
\usepackage{url}            
\usepackage{booktabs}       
\usepackage{amsfonts}       
\usepackage{nicefrac}       
\usepackage{microtype}      
\usepackage{xcolor}         

\usepackage{color, colortbl}
\usepackage{multirow}
\usepackage{lscape}
\usepackage{graphicx}
\usepackage{amsmath,amssymb}
\usepackage{todo}

\setcitestyle{numbers,square,comma}
\include{notation}

\newcommand{\ghls}{\gG_{\texttt{hls}}}
\newcommand{\mlp}{\texttt{MLP}}
\newcommand{\coolname}{\texttt{gnn4hls}}
\definecolor{Gray}{gray}{0.85}

\title{A Graph Deep Learning Framework for High-Level Synthesis Design Space Exploration}

\author{%
  Lorenzo Ferretti${}^1$\thanks{These two authors contributed equally. Corresponding author: ferrelo@cs.ucla.edu}, Andrea Cini${}^{2}$\footnotemark[1], Georgios Zacharopoulos${}^{3}$, Cesare Alippi${}^{2,4}$, Laura Pozzi${}^{2}$\\
  ${}^1$University of California Los Angeles, ${}^2$Universit\`a della Svizzera italiana, \\ ${}^3$ Hardvard University, ${}^4$ Politecnico di Milano\\
}

\begin{document}

\maketitle

\begin{abstract}
The design of efficient hardware accelerators for high-throughput data-processing applications, \eg\ deep neural networks, is a challenging task in computer architecture design. 
In this regard, High-Level Synthesis~(HLS) emerges as a solution for fast prototyping application-specific hardware starting from a C/\CC\ behavioural description of the application computational flow. In the accelerator synthesis phase, designers apply HLS directives to optimize the hardware implementation, by trading-off cost and performance. 
This Design-Space Exploration~(DSE) aims at identifying Pareto optimal synthesis configurations whose exhaustive search is often unfeasible due to the design-space dimensionality and the prohibitive computational cost of the synthesis process.
Within this framework, we effectively and efficiently address the design problem by proposing, for the first time in the literature, graph neural networks that jointly predict acceleration performance and hardware costs of a synthesized behavioral specification given optimization directives. The learned model can be used to rapidly approach the Pareto curve by guiding the DSE, taking into account performance and cost estimates. The proposed method outperforms traditional HLS-driven DSE approaches, by accounting for arbitrary length of computer programs and the invariant properties of the input.
We propose a novel hybrid control and data flow graph representation that enables training the graph neural network on specifications of different hardware accelerators; the methodology naturally transfers to unseen data-processing applications too.
Moreover, we show that our approach achieves prediction accuracy comparable with that of commonly used simulators without having access to analytical models of the HLS compiler and the target FPGA, while being orders of magnitude faster. Finally,  the learned representation can be exploited for DSE in unexplored configuration spaces by fine-tuning on a small number of samples from the new target domain. The outcome of the empirical evaluation of this transfer learning shows strong results against state-of-the-art baselines in relevant benchmarks including neural processing.
\end{abstract}

\input{content}




\bibliography{main.bib}

\clearpage

\appendix

\input{appendix}



\end{document}

%% file: notation.tex
\usepackage{bm}

\newcommand{\newterm}[1]{{\emph{#1}}}

\def\eqref#1{equation~\ref{#1}}

\def\1{\bm{1}}

\newcommand{\ie}{\textnormal{i.e.,}}
\newcommand{\eg}{\textnormal{e.g.,}}
\newcommand{\wrt}{\textnormal{w.r.t.}}

\def\vc{{\bm{c}}}

\def\ve{{\bm{e}}}

\def\vs{{\bm{s}}}

\def\vu{{\bm{u}}}
\def\vv{{\bm{v}}}

\def\mE{{\bm{E}}}

\def\mV{{\bm{V}}}

\DeclareMathAlphabet{\mathsfit}{\encodingdefault}{\sfdefault}{m}{sl}
\SetMathAlphabet{\mathsfit}{bold}{\encodingdefault}{\sfdefault}{bx}{n}

\def\gG{{\mathcal{G}}}

\def\sR{{\mathbb{R}}}

\newcommand{\tuple}[1]{{\left\langle #1 \right\rangle}}

\newcommand{\concat}{\mathbin{\|}}

\newcommand{\CC}{C\nolinebreak\hspace{-.05em}\raisebox{.4ex}{\tiny\bf +}\nolinebreak\hspace{-.10em}\raisebox{.4ex}{\tiny\bf +}}
\def\CC{{C\nolinebreak[4]\hspace{-.05em}\raisebox{.4ex}{\tiny\bf ++}}}

%% file: content.tex
\section{Introduction}

In the last decade, specialised hardware has become a viable solution to address the end of Moors’s Law~\citep{MooreLaw} and the breakdown of Dennard scaling~\citep{DennardScaling} and deal with the constant growth in performance and efficiency requirements of computer systems.
Whitin this context, short time to market, higher design productivity and reusability of existing modules are only a subset of the challenges that have driven the Electronic Design Automation~(EDA) industry and research.
In particular, the recent wide-spread of solutions based on artificial intelligence and deep learning techniques~\citep{lecun2015deep, schmidhuber2015deep} has lead to an increasing demand of hardware accelerators able to target a large variety of computational workloads. 

Automating the design process by exploiting the predictive power of modern machine learning (ML) models is an appealing approach that, while accelerating the development of computer architectures, would also allow the ML community to benefit from the improved computing platforms. In fact, progress in one field ripples through the other one, thus creating a positive feedback loop and a virtuous cycle~\citep{fahim2021hls4ml}. In this setting, graphs are natural candidates to capture the functionally-dependant structure of software and hardware systems. It is not surprising, then, that the advent of graph neural networks~(GNNs)~\citep{scarselli2008graph, bronstein2017geometric, battaglia2018relational} led to impressive developments in computer-aided hardware and software design: from chip placement~\citep{mirhoseini2020chip} to compilers for ML computational graphs~\citep{zhou2020transferable}.
In this context, we advocate for the adoption of analogous automation strategies to fill the gap between the increasing request for efficient and effective hardware acceleration of existing applications and the hardware design productivity.

\begin{figure}[t]
    \centering
    \includegraphics[width=\linewidth]{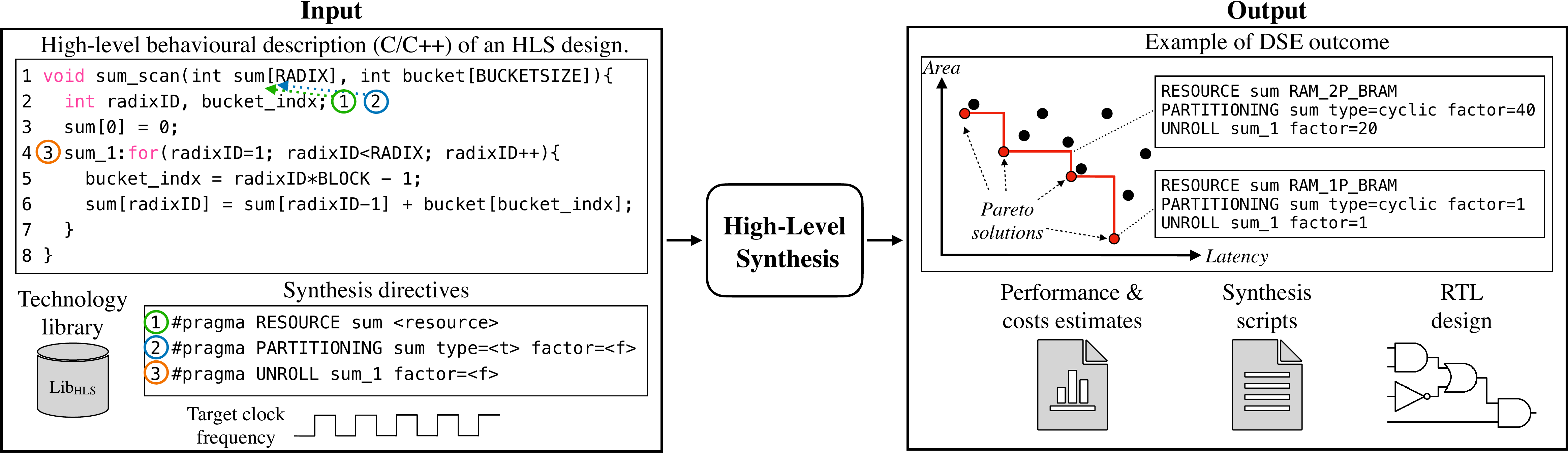}
    \caption{Example of HLS design flow. A behavioural description, synthesis directives, technology library, and a target frequency are given to the HLS tool, which generates as a result an RTL design, performance and cost reports, as well as the synthesis scripts.}
    \label{fig:hls_process}
    \vskip -1em
\end{figure}

Traditional Integrated Circuits~(ICs) design methodologies rely on Hardware Description Languages~(HDLs) to describe logical components and their interaction at a Register Transfer Level~(RTL). This approach requires designers to manually define the concurrent description of millions of transistors working in parallel to carry out the desired computations. To reduce the burden of this task, High-Level Synthesis~(HLS) comes into play.
HLS tools enable to start the design process from high-level behavioural specifications in C/\CC/SystemC, hence avoiding designers to deal with the tedious and error prone task of implementing the functionality at RTL level.
Besides specifying the desired behavior, as shown in~Figure~\ref{fig:hls_process}, designers can guide the synthesis process by applying directives able to tune the resulting RTL implementation according to target performance and cost requirements. Synthesis directives allow to specify how to implement in hardware specific software constructs such as loops, arrays and functions. For example, the designer can use directives to tune the degree of hardware parallelization of a loop by specifying a loop unrolling factor. 
While HLS allows to explore a vast design space of micro-architectural variations by using different directives, resulting performance and resource utilization of each implementation cannot be determined a priori. In fact, exhaustive exploration involves time-consuming syntheses, whose number grows exponentially \wrt\ the number of applied optimizations. Moreover, among all the possible configurations~(\ie\ combinations of directives), only a few are Pareto-optimal from a performance and costs perspective. 
As designers are interested in effective methodologies to automate the Design-Space Exploration~(DSE) process, the HLS-driven DSE problem consists in identifying as accurately as possible the set of Pareto implementations while, at the same time, minimizing the number of synthesis runs. 

Recent works demonstrated the possibility to guide DSEs by exploiting the notion of function similarity and the knowledge acquired from past explorations performed on different functions~\citep{ferretti2020leveraging, jhye2020transfer, WangJun20} and have shown strong empirical results besides the small number of source domains. In light of this, \textit{we claim that data-driven approaches are the way forward for HLS-driven DSE}. 
In this work, we introduce the methodological framework where we firmly believe that progress in the field is bound to happen. We propose both a data representation able to capture the critical elements of the HLS process and the data-processing tools to profit from such representation. 
At first, we introduce a novel graph representation of computer programs, based on an augmented hybrid control and data flow graph, capturing invariant properties and relevant information from the HLS perspective.
Then, we exploit this representation to train a novel graph neural network model in a supervised fashion, by fitting the model on a dataset of previously synthesized configurations (behavioral specifications plus optimization directives) to predict the latency and resource utilization corresponding to each point. To the best of our knowledge, this is the first attempt to use graph representation learning from software specifications to perform HLS-driven design space exploration. We show that the learned model can be used for effective DSE after fine-tuning on a small set of samples and  that our method compares favorably against state-of-the-art baselines on relevant benchmarks. We refer to our framework as \coolname. We believe that the results achieved here constitute a strong signal for the community that calls for a general effort in collecting large datasets of syntheses to unlock the full potential of graph deep learning solutions to the HLS-driven DSE problem. In order to accellerate progress in this direction, methods and datasets presented here together with a platform for data collection will be open sourced at the end of the blind review process.

The rest of the paper is organized as follows. In Section~\ref{s:background} we define the problem by introducing the main concepts and proper terminology for HLS-driven DSE and graph neural networks. Then, in Section~\ref{s:methods} we lay out the details of our approach. In Section~\ref{sec:experimental_evaluation} we evaluate the proposed method on relevant benchmarks. Finally, we discuss the related works in Section~~\ref{sec:related_work}, draw our conclusions and discuss future works in Section~\ref{s:conclusion}.

\section{Background}\label{s:background}

\subsection{High-Level Synthesis driven Design-Space Exploration}

Given a software functionality, \eg\ the \texttt{sum\_scan} function from the Radix Sort algorithm in Machsuite~\citep{reagen2014machsuite} ~(used as a running example in the rest of the document), we define as \newterm{HLS design}, or simply \newterm{design}, the functionality to be realized in hardware, and as \newterm{specifications} the behavioural description of the design in a high-level programming language such as C/C++. The specification~($SW$) is given in input to the HLS tool together with the target technology library~($Lib_{HLS}$) and a target frequency~($F$).
The result of the synthesis process is named \emph{implementation}, and it is an automatically generated RTL code, usually in VHDL or Verilog. The resulting RTL is coupled to a performance metric~(\eg\ latency or throughput), and a cost metric~(\eg\ area or energy costs).

An implementation is generated by applying a set of \emph{directives} -- specified using compiler pragmas -- to the $SW$, affecting the resulting performance and costs. The set of directives affecting an implementation is named \emph{configuration}.
Each directive is associated to a \emph{target} in the $SW$, which can be either a label or a code construct (e.g., labeled statements, function names, variables, etc.). In addition, a directive is characterized by its \emph{type} and an associated \emph{value}. Examples of directive types are \texttt{loop unrolling} -- affecting number of resources required to implement the loop body in hardware and enabling its parallel execution -- and \texttt{array partitioning} -- splitting the input array in multiple memory banks and enabling parallel access to the data. 
The directive value forces a given directive type to a specific value. As an example, an unrolling factor of $2$ doubles the logic required to implement in hardware the loop body, enabling parallel hardware execution of two iterations.

The HLS design flow and the different elements characterizing it are shown in  Figure~\ref{fig:hls_process}. Figure \ref{fig:hls_process}(left) shows the inputs of the HLS design flow: a behavioural description of an HLS-design -- the \texttt{sum\_scan} function in the Radix Sort benchmark from MachSuite~\citep{reagen2014machsuite} -- an example of \texttt{pragmas} applied to the array and loop constructs of the specification, the technology library adopted, and the target frequency. After being processed by the HLS tool, the resulting implementation is generated as an RTL desing, with the synthesis scripts and the performance and cost reports--Figure~\ref{fig:hls_process}(right).

Given a \emph{design} $D$, a designer limits the set configurations to explore during a DSE by defining a \emph{configuration space}.
The \emph{configuration space} $X_D$ is defined as the Cartesian product among the set of \emph{directive values} $V_i$ associated to each $i$-th \emph{directive}, \ie\ $X_D = V_1 \times V_2 \times ... \times V_N$, where $N$ is the number of considered \emph{directives}. The size of the \emph{configuration space} is given by its cardinality $|X_D|$.
Given a configuration space $X_D$, a \emph{design space} $Y_D$ can be defined as the set of \emph{implementations} resulting from the synthesis of the configuration in $X_D$.

\paragraph{Task formulation} The DSE problem is a Multi-Objective Optimization Problem (MOOP) having costs and merit as objective functions.
In the context of hardware design common performance measure and cost are  \emph{latency}, and \emph{area} or \emph{power} respectively. In this work we use as performance the \emph{effective latency}~(LAT), \ie\ the number of clock cycles required by the hardware implementation to execute its functionality multiplied by the target clock of the system.
For cost we consider the percentage of resources~(silicon) utilization required to implement the IC in hardware. In this work, since our target architecture is a Field Programmable Gate Array (FPGA), costs are expressed in terms of number of Flip-Flops (FF), Look-Up Tables (LUT), Digital Signal Processor (DSP), and Block RAM (BRAM)\footnote{For the HLS designs considered in this work, the DSEs have not affected the number of BRAM; therefore, we have not include BRAM estimation in our experiments.}.
In particular, the objective is to identify a subset $P_D$ of the configuration space $X_D$ such that: $P_D = \{x | x \in X_D$ and $x$ is Pareto$\}$.
A \emph{Pareto configuration} ($p$) of a design $D$ is defined as: $p$ $\in$ $P$ $\Leftrightarrow$ $\nexists$ $x$ $\in$ $X_D$, $x$ $\neq$ $p$ $|$ $a_x$ $\leq$ $a_p$ $\wedge$ $l_x$ $\leq$ $l_p$. With $a_p$, $a_x$, and $l_p$, $l_x$ being the cost ($a$) and merit ($l$) associated to the implementation of $p$ and $x$ respectively.

\subsection{Graph neural networks}

Input behavioural specifications~(programs) significantly differ in terms of size, structure, constructs, and optimization directives. This variability is hardly captured by vector representations which have been, we argue, one of the main limiting factors of previous works in attempting to learn predictive models of the HLS process~\citep{LiuJun13, WangJun20}. Differently, when modeling the problem in the space of graphs we can use methods that naturally exploit existing functional dependencies, account for the variability in the structure of different specifications, and seamlessly transfer learned representations across different configuration spaces. Furthermore, graph-processing techniques permit to exploit the properties of graph representations (\eg\ permutation invariance) inducing positive inductive biases that restrict the hypothesis space explored by the learning system to plausible models.

We consider attributed directed graphs, with attributes associated to both nodes and edges. In particular, a graph $\gG$ is a tuple $\tuple{\mV, \mE, \vu}$, where $\mV = \{ 1, \dots, N\}$ is a set of nodes, $\mE = \{(i,j) \vert i,j\in \mV\}$ is a set of edges and $\vu \in \sR^u$ is a global attribute vector. We denote by $\vv_i \in \sR^v$ the raw features associated with node $i \in \mV$ and with $\ve_{i,j} \in \sR^e$ the attribute vector associated with edge $(i,j) \in \mE$ connecting nodes $i$ and $j$.

Different works propose general frameworks to design GNNs: inspired by~\citet{gilmer2017neural} and \citet{battaglia2018relational},  we consider a very general class of message-passing neural networks~(MPNNs) with global attributes where the $t$-th propagation layer~(or step) can be written as
\begin{align}\label{eq:mp}
    \vv_i^t &= \tau^t_\vv\left(\vv_i^{t-1}, \texttt{AGGR}_\vv\Big\{ \psi^t_\vv\left( \vv_j^{t-1}, \vv_i^{t-1}, \ve_{j,i}\right); {(j,i) \in \mE}\Big\}, \vu^{t-1} \right)\\
    \vu^t &= \tau_\vu^t \left(\vu^{t-1}, \texttt{AGGR}_\vu\Big\{\psi^t_\vu\left(\vv_i^t, \vu^{t-1}\right); i \in \mV\Big\}\right),
\end{align}
where the update ($\tau^t_\vv$,$\tau^t_\vu$) and message~($\psi^t_\vv$,$\psi^t_\vu$) functions can be implemented by any differentiable function, \eg\ Multi-Layer Perceptrons~(MLPs). Aggregation functions~$\texttt{AGGR}_\vu\{{}\cdot{}\}$ and $\texttt{AGGR}_\vv\{{}\cdot{}\}$ can be any permutation invariant operation. Multi-layer GNNs are built by stacking several propagation layers, which allow to aggregate at each node messages from different neighborhoods.
\section{Methods}\label{s:methods}

\subsection{Data representation}

\begin{figure}[t]
    \centering
    \includegraphics[width=\textwidth]{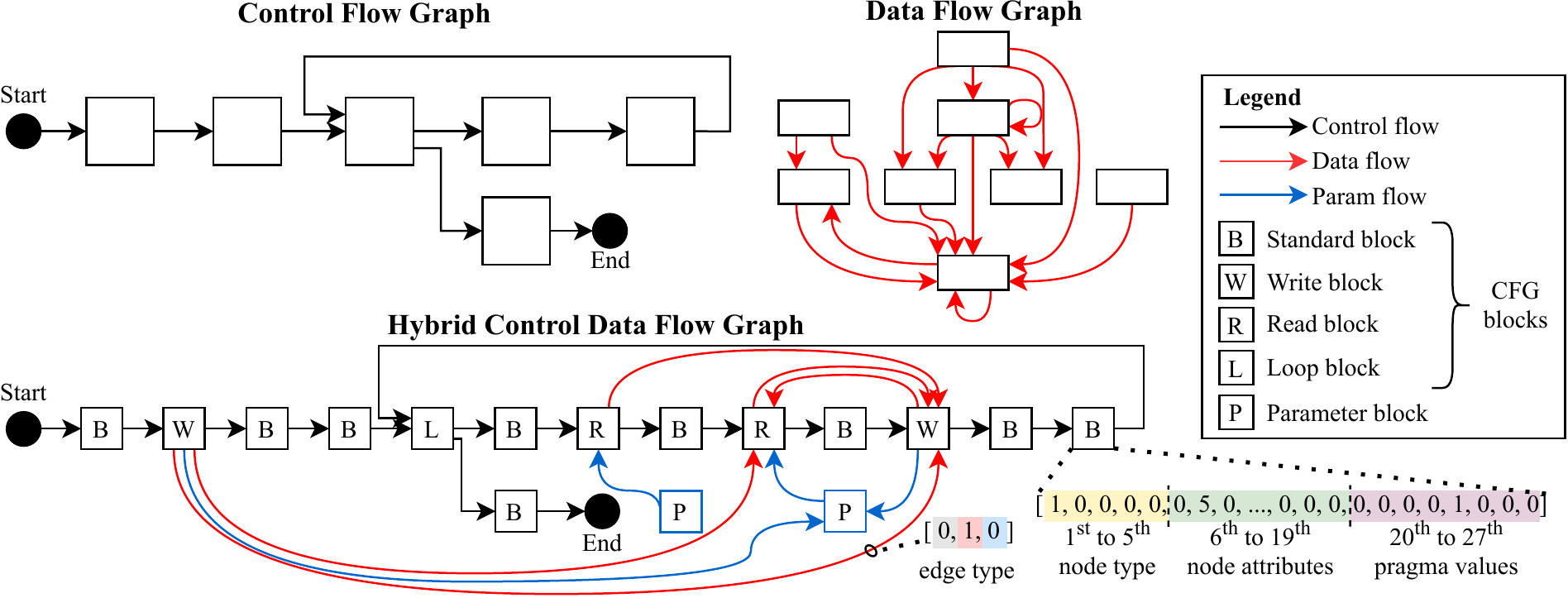}
     \vskip -0.5em
    \caption{Data representation. Control Flow Graph, Data flow graph and Hybrid Control Data Flow Graph representations of the behavioural specification of the function in Figure~\ref{fig:hls_process}.}
    \label{fig:graphs_representations}
    \vskip -1em
\end{figure}

Control Flow Graphs (CFGs) are graphs representing the possible execution paths in a program. 
The CFG is a directed graph $\gG_\texttt{CFG}$ defined as the tuple $\tuple{\mV_\texttt{CFG}, \mE_\texttt{CFG}}$, where $\mV_\texttt{CFG}$ is the the set of nodes corresponding to the basic blocks of the program, and $\mE_\texttt{CFG}$ is the set of edges representing the possible control flows among basic blocks. An example of CFG for the \texttt{sum\_scan} function is shown in Figure~\ref{fig:graphs_representations}-(top).
Traditionally, a basic block is defined as a consecutive sequence of instructions without incoming and outgoing branches except for the first and last instructions of the block respectively.
In this work we adopt a different definition of basic block. In particular, in our proposed formulation, we differentiate basic blocks according to the type of instructions they perform. We discriminate between the following types of blocks: \emph{loop blocks} identifying basic blocks including loop instructions, \emph{read block} including a single load instruction from main memory, \emph{writes block} including a single store instruction to main memory, \emph{function block} including a function invocation instruction, and lastly \emph{standard block} being basic block including instructions performing computations that do not belong to any of the above mentioned categories.
The effect of this representation and taxonomy affects the granularity of the CFG representation, increasing the number of blocks \wrt the traditional one. Figure \ref{fig:graphs_representations}-(bottom), shows the differences of the proposed CFG representation -- with a higher block granularity -- w.r.t. the traditional one.  This choice aims at avoiding the limitation of approaches relying on a vector-based representation of the program and directives~\citep{LiuJun13}, while, at the same time, focusing only on the information that is more relevant from the HLS and DSE perspective. Notably,  recent works highlight the effectiveness of adopting a similar taxonomy of basic blocks to capture similarities among points in different configuration spaces~\citep{ferretti2020leveraging}. Compared to previous works in program analysis~(\eg~\citet{li2019graph}), we include in our CFG representation only corse-grained information on the types of operations performed in each block.
In addition to the node type, attribute vectors associated to each node include block-type related information, such as: number of instruction in a block, number of iterations of a loop block, presence of loop carried dependencies, and other type of information extracted through static and dynamic code analysis performed using custom LLVM \citep{lattner2004llvm} compiler passes~(more details in Section~\ref{sec:experimental_evaluation}).
While CFGs contain information about the execution flow of a program, they do not model the flow of data and information. Data Flow Graphs~(DFGs) address this aspect.
DFGs are used to represent the different dependencies between the instructions of a program. In particular they represent the use-def chains among variables in the program execution. The DFG is a directed graph $\gG_\texttt{DFG}$ defined as a tuple $\tuple{\mV_\texttt{DFG}, \mE_\texttt{DFG}}$, where $\mV_\texttt{DFG}$ is the a set of nodes corresponding to a instruction in the input source code, and $\mE_\texttt{DFG}$ is the set of edges representing the possible data flow among instruction. The DFG representation for the running example function is shown in Figure~\ref{fig:graphs_representations}-(top).

\paragraph{Hybrid Control Data Flow Graph}

HLS tools use instrumented Control Flow Graphs and Data Flow Graphs as program representations to decide how to implement in hardware the design functionality. In our approach we aim at using a similar representation directly as input of a learning method. We propose a graph representation of the software description including both CFG and the DFG information. In particular, we augment the CFG representation by adding data flow edges and nodes representing the input and output parameters of the function. Data flow edges are added among the nodes involving operations affecting the input and output parameters. These edges are identified by tracking the def-use chains among parameter variables in the DFG and embedding them in the CFG. In addition, edges among the parameter nodes and the CFG read and write blocks are added to the set of edges (\textit{param flow}). We indicate this \textit{Hybrid Control Data Flow Graph} as $\ghls$ and we use it as input representation in our methodology. To sum up, $\ghls$ is defined as a tuple $\tuple{\mV_{\texttt{hls}}, \mE_{\texttt{hls}}, \vu}$, where $\mV_{\texttt{hls}}$ is the a set of nodes corresponding to basic blocks and function parameters, and $\mE_{\texttt{hls}}$ is the set of attributed edges representing the control, data, and parameter flows. Each edge attribute vector $\ve_{i,j}$ is a $3$ dimensional feature vector with a one hot encoding representation of the edge type. Lastly, $\vv_i \in \mV_{\texttt{hls}}$ is the feature vector associated to each node with a one-hot-encoding representation of the nodes type and their attributes, plus the value of each optimization directive.
Figure \ref{fig:graphs_representations}-(bottom) shows the  Hybrid Control Data Flow Graph representations for the running example function.

\subsection{Graph neural networks for high-level synthesis design-space exploration}
\begin{figure}[t]
    \centering
    \includegraphics[width=\textwidth]{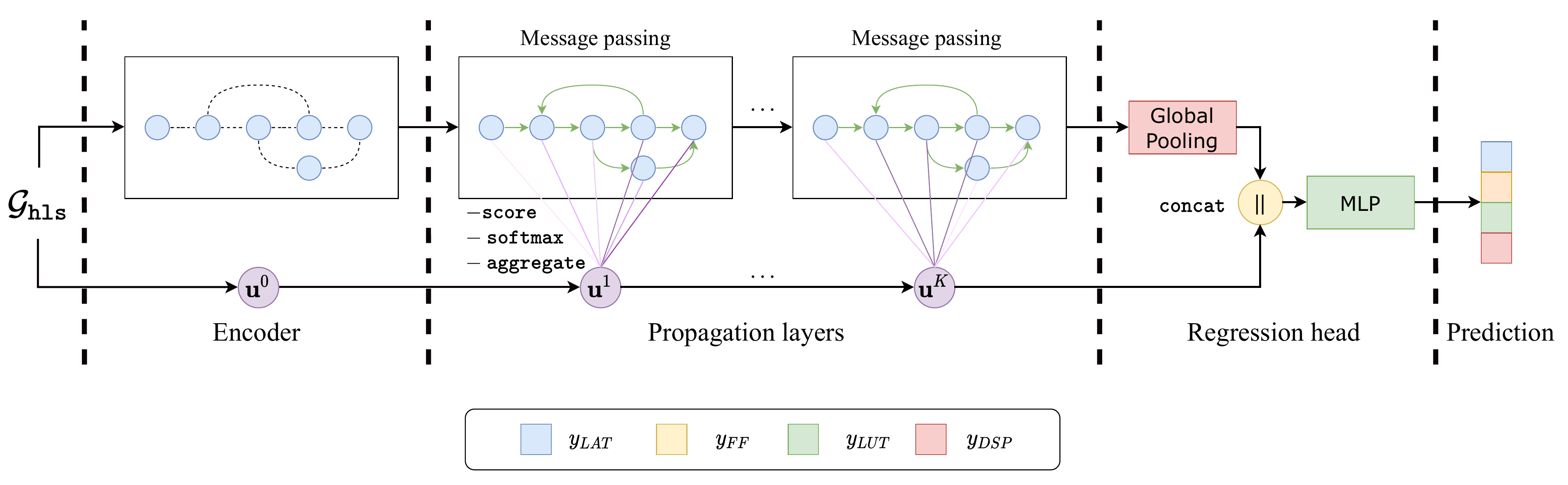}
     \vskip -0.5em
    \caption{\texttt{gnn4hls}. An encoder block maps the pre-process input representation with standard MLPs. Propagation layers perform message passing and update the global representation with multi-head attention layers. The final block, maps the latent representation learned by the network into the regression targets.}
    \label{fig:gnn4hls}
    \vskip -1em
\end{figure}

Given a graph $\ghls$ representing a program annotated with optimization directives, we process it with message passing neural networks by parametrizing the computation of messages exchanged between neighboring nodes of the graph with MLPs. A schematic of the model is shown in Figure~\ref{fig:gnn4hls}: we describe at first each computational block in detail then discuss the training procedure.

\paragraph{Encoder} The Encoder block maps node, edge, and global graph features into a first hidden representation without performing any message-passing operation. The Encoder is implemented by using standard MLPs as update functions
\begin{equation*}
\begin{split}
    \vv^0_i = \mlp_{\vv}^{\texttt{enc}}\left(\vv_i\right),
\end{split}\quad
\begin{split}
    \vu^0 = \mlp_{\vu}^{\texttt{enc}}\left(\vu\right),
\end{split}\quad
\begin{split}
    \ve^{\texttt{enc}}_{i,j} = \mlp_{\ve}^{\texttt{enc}}\left(\ve_{i,j}\right),
\end{split}
\end{equation*}
as using feed-forward layers before message passing has shown to be beneficial to final performance in GNNs~\cite{you2020design}. The Encoder block is followed by a stack of propagation layers.

\paragraph{Propagation layer} Propagation layers are instantiated in the message-passing framework shown in Equation~\ref{eq:mp}. In particular, the node update and message functions are implemented as:
\begin{equation}
\vv_i^t = {\mlp}^t_{\tau_\vv}\bigg(\vv_i^{t-1} \concat\ \texttt{MEAN}\Big\{ {\mlp}^t_{\psi_\vv}\big( \vv_j^{t-1} \concat\ \ve^{\texttt{enc}}_{j,i}\big); (j,i) \in \mE_{\texttt{hls}}\Big\}\bigg),
\end{equation}
where $\concat$ is the vector concatenation operator and $\texttt{MEAN}\{{}\cdot{}\}$ indicates that we aggregate incoming messages by averaging them out. To update the global representation $\vu^t$ we use a MLP as update function that takes in input the concatenation of $\vu$ at the previous propagation step and node attributes aggregated by exploiting the attention mechanism~\citep{vaswani2017attention, velickovic2018graph}. In particular, we use a MLP~($\mlp_\alpha({}\cdot{})$) to compute a raw attention score for each node attribute vector $\vv_i^t$ given the global features $\vu^{t-1}$; raw scores~(\ie\ logits) are then normalized over the nodes of the graph with a \texttt{softmax} function. Normalized node scores are used to aggregate node features  processed by a third MLP. Putting all together, each propagation layer updates the global representation as follows:
\begin{equation}
\begin{gathered}
\texttt{score}\left(\vu, \vv_i\right) = \texttt{softmax}\left\{\mlp_\alpha\Big(\vu, \vv_i\Big); i \in \mV_{\texttt{hls}}\right\} \\
\vu^t = {\mlp}^t_{\tau_\vu}\left(\vu^{t-1} \concat\ \texttt{SUM}\Big\{ \texttt{score}\left(\vu^{t-1}, \vv^t_i\right) \odot {\mlp}^t_{\psi_\vu}\big( \vv_i^{t-1}\big); i \in \mV_{\texttt{hls}}\Big\}\right),
\end{gathered}
\end{equation}
where ${}\odot{}$ is the element-wise multiplication operator and $\texttt{SUM}\{{}\cdot{}\}$ indicates aggregation by graph-wise summation of node features. In practice multiple attention heads can be used in parallel for increased model capacity.

\paragraph{Regression head} After $T$ message-passing blocks, node representations are pooled in a single vector using a permutation invariant aggregation function~(\eg\ by taking the sum or the average of node representations, optionally weighted by learned attention scores). The pooled representation is then concatenated to the global attributes $\vu^T$ leading to a vector representation of the input graph. This feature vector is fed trough a last MLP which maps it to a prediction of latency and resources as shown in Figure~\ref{fig:gnn4hls}.

\paragraph{Training procedure and transfer learning} We train the GNN by supervised learning to predict the outcome of the HLS procedure. We use data from several synthesized designs, with program specifications relevant to different domains~(we refer to Section~\ref{sec:experimental_evaluation} for more details). While learning to predict the outcome of the synthesis process for configurations spaces already partially explored is interesting from a research perspective, we are interested in assessing the possibility of exploiting the model for DSE. In particular, we aim at assessing if the learned representation can support transfer to different domains~(designs) when only a few samples, or none, from the target configuration space are available. We comment that our method differs from previous approaches, which usually are domain specific and tied to the characteristics of the target design space. Instead, our methodology is general and can easily incorporate knowledge from different design spaces by simply including synthesized points in the training dataset.

\section{Experimental evaluation}\label{sec:experimental_evaluation}

The graph representation of HLS designs and the associated \texttt{pragma} values are generated combining LLVM~\citep{lattner2004llvm} compiler passes, Clang Abstract Syntax Tree (AST) analysis, Frama-C~\citep{Couq2021framac} internal representation of program dependencies, and HLS synthesis information from a recently published database of HLS-drived DSEs~\citep{ferretti2021db4hls}. The custom LLVM pass generates the CFG representation from the compiler Intermediate Representation (IR) and performs static program analysis to identify the block proprieties. 
In order to account for the information lost in the LLVM IR representation, 
an AST visitor extracts and maps the SW information to the CFG blocks. Data flow information is extracted from the Frama-C program dependency analysis, and used to generate the data flow edges of $\ghls$. 
We generated graph representations from $23$ different functions in MachSuite~\citep{reagen2014machsuite}. 
The considered functions include a wide range of computational intensive applications such as: matrix-matrix multiplication, sparse matrix-vector multiplication, sorting algorithms, stencil computations, molecular dynamics, and forward passes of fully-connected neural networks.
For each design, we used configurations and synthesis results available from \texttt{db4hls}~\citep{ferretti2021db4hls}, an open source data base of HLS DSEs. The total number of configurations considered in this work is $103093$. Configuration spaces contain from several hundreds, up to several thousand design points. 
For the graph neural network, we use as global graph attributes the number of LLVM instructions, the number of input parameters, and the average value of each directive set within the configuration minus the mean value of the directive sets computed over the entire configuration space. To increase robustness to outliers, we also concatenate to the representation the same values minus the median.

In the following at first we perform an experiment to asses the accuracy of our model in inferring the performance and costs of unseen directive combinations, then we switch to the DSE settings. Hyper parameters of the models and full experimental setup are provided in the supplementary materials.

\subsection{Performance and cost estimation}

\begin{figure*}[t]
\begin{minipage}[t]{0.32\textwidth}
    \centering
    \includegraphics[width=\linewidth]{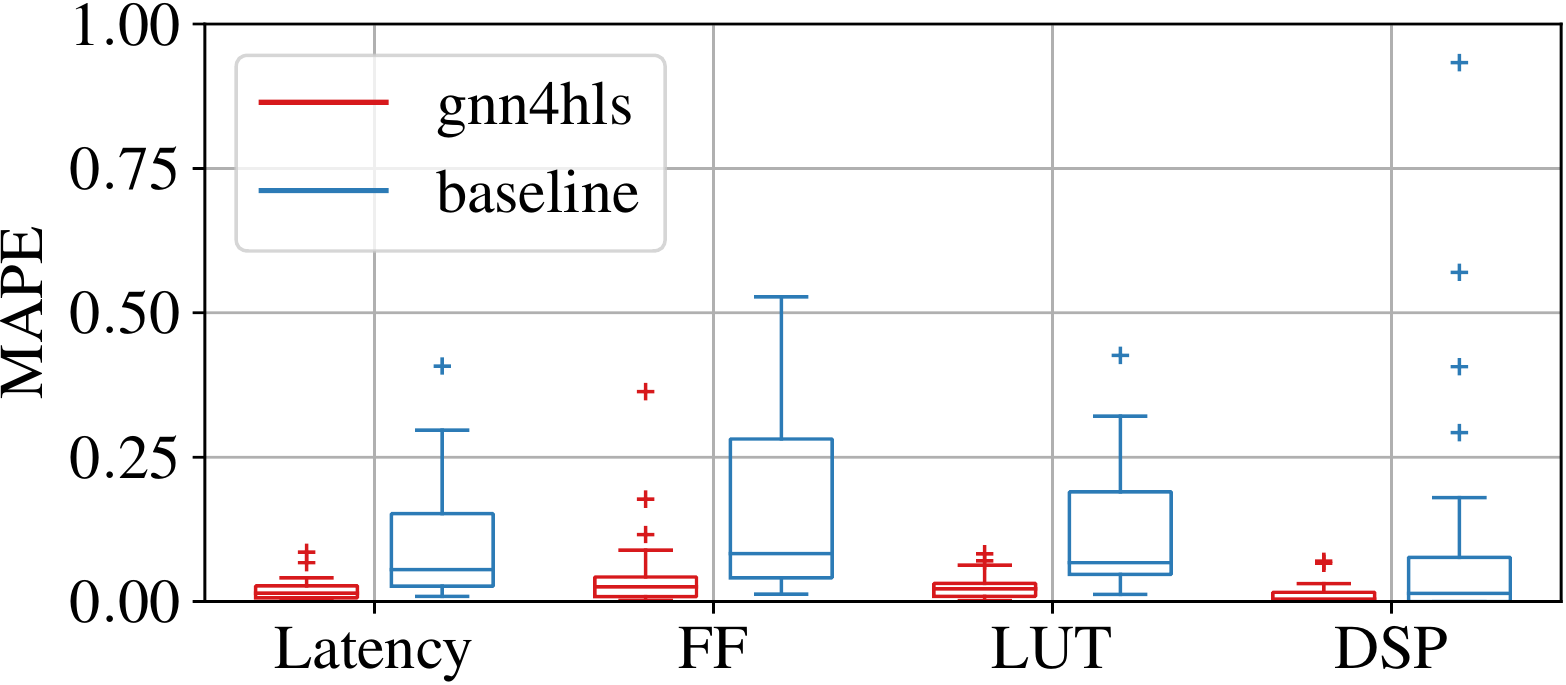}
     \vskip -1em
    \caption{Comparison among the \texttt{gnn4hls} approach proposed in this work and a MLP ones adopted as a baseline.}
    \label{fig:comparison}
\end{minipage}%
\hfill
\begin{minipage}[t]{0.32\textwidth}
    \centering
    \includegraphics[width=\linewidth]{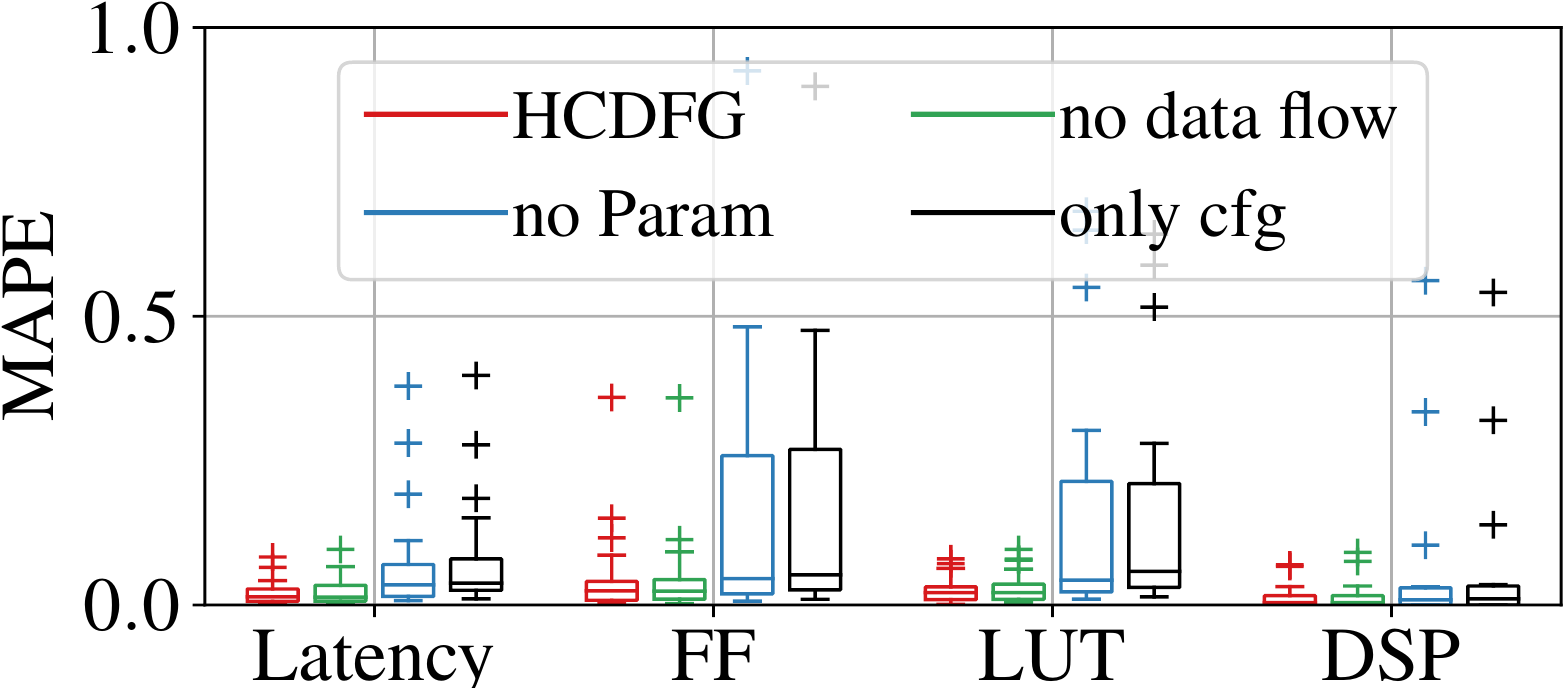}
     \vskip -1em
    \caption{Ablation study of \coolname~graph structure.}
    \label{fig:ablation}
\end{minipage}%
\hfill
\begin{minipage}[t]{0.32\textwidth}
  \centering
    \includegraphics[width=\linewidth]{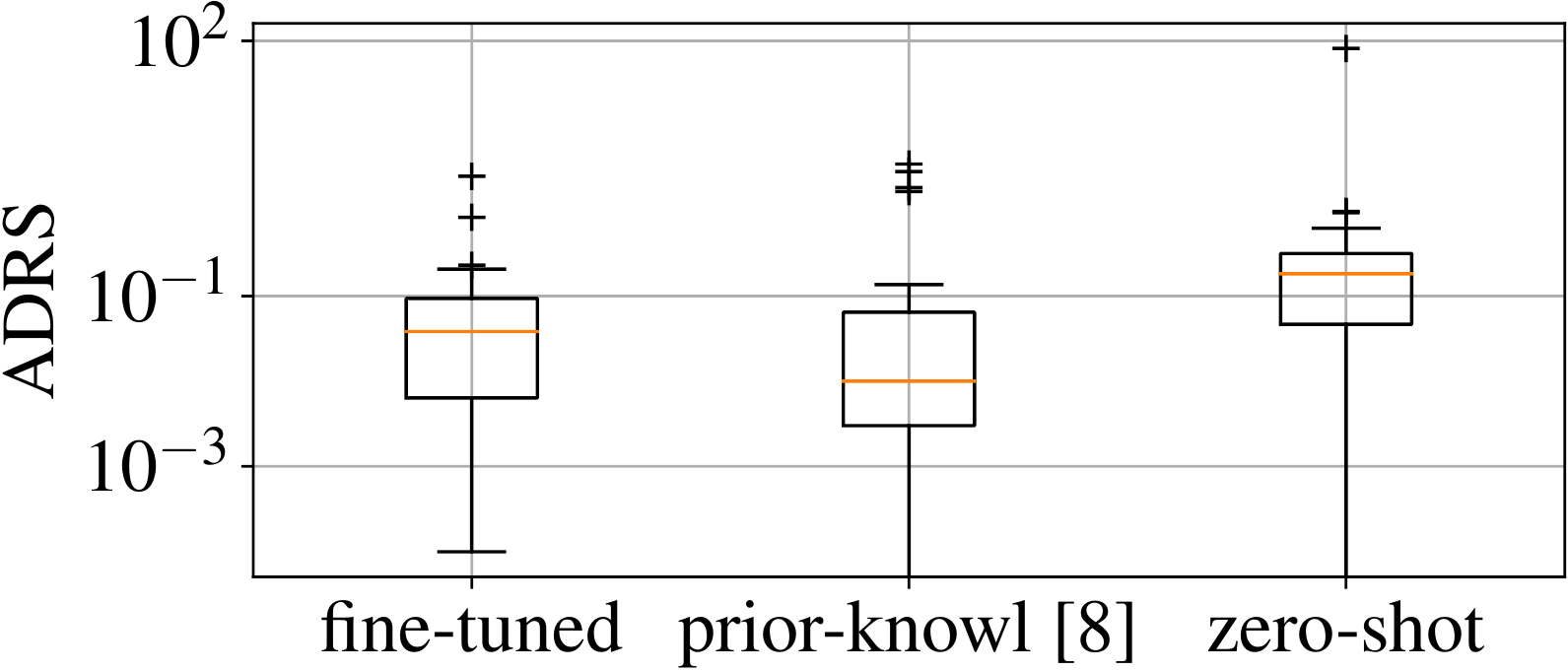}
     \vskip -1em
    \caption{Comparison between fine-tuned, prior-knowledge~\citep{ferretti2020leveraging}, and zero-shot approaches.}
    \label{fig:approaches_comparison}
\end{minipage}
\vskip -1.5em
\end{figure*}

We split the synthesized configuration space of each function in three folds, keeping $70\%$ of the available points for training, $10\%$ for validation, and $20\%$ for testing. Selecting a proper baseline for comparison is not easy since none of the approaches existing in the literature can easily be extended to our settings. The most similar approach is the one introduced by~\citet{jhye2020transfer}, where a MLP is trained in a multi-domain setting. However their approach, based on multi-task learning~\citep{caruana1997multitask}, relies on training different input and output layers for each domain, hence limiting flexibility. Furthermore, they use only optimization directives as an input and predict normalized scores for performance and costs instead of the actual latency and resources. For these reasons we consider a DeepSets~\cite{zaheer2017deep} model to be a more appropriate baseline: in practice we use a node-level MLP followed by a permutation invariant aggregation and a second MLP to process the aggregated features. The network architectures were chosen to have a similar model complexity.
The boxplot in Fig.~\ref{fig:comparison} shows, for each figure of merit, the median, $1^{st}$ and $3^{rd}$ quartiles, and interquartile range of the mean absolute percentage error (MAPE) of the two models over the 23 functions in the dataset. Results, averaged over $5$ independent runs, show that our model drastically outperforms the baseline by achieving an average MAPE, averaged over performance and costs estimates, of $2.7\%$ against $15.4\%$: an improvement of over $82\%$. 
Furthermore, the performance obtained by our model is qualitatively similar to SoA simulator ones, which exploit an analytical model of the HLS process, HLScope+~\citep{choi2017hlscope+} for the latency, and MPSeeker~\citep{zhong2017design} for resources\footnote{A direct comparison with these methodologies was not possible since these were not available open source. Thus, we compare \wrt~ the performance of the original papers.}. 
In particular, latency estimation performance are comparable to the best from the SoA (1.1\% MAPE of HLScope+ vs. 2.1\% of \coolname), our area performance estimation outperforms the ones from existing models. \coolname~ achieves 4.8\%, 2.6\% and 1.3\% estimation for FFs, LUTs, and DSPs compared to the 14.7\%, 13.2\%, 12.7\% respectively from MPSeeker.
In addition, inference time is greatly reduced from seconds to tens of milliseconds \wrt~SoA alternatives. In particular, the network used here requires $\approx20\text{ms}$ to process a single point from the \texttt{get\_delta\_matrix\_weights1} function on an Intel(R) Xeon(R) Silver CPU.

\subsection{Ablation study} To evaluate the effectiveness of the proposed graph representation, we have performed an ablation study on the graph edges. Fig.~\ref{fig:ablation} shows results obtained given 3 different ablations on the graph structure: no data flow edges, no param edges, and both data flow and param edges removed. For all the representation the same type of nodes and attributes have been considered. Results show that the proposed Hybrid CDFG representation leads to the best results. Note that the DFG can largely be inferred from Param edges, thus the similar performance of this setting.

\subsection{Design-space exploration}\label{sec:experimental_evaluation_dse}

The second set of experiments aims at addressing DSE. Herein, we focus on approximating the Pareto-frontier of a target HLS design, given the synthesis outcomes of all the considered functions but the target one~(\ie\ we perform a leave-one-out evaluation \wrt\ the available functions). 
We consider the setting where the designer performs an initial na\"ive random sampling of the configuration space, uses the synthesized points to fine tune the model and then uses the model's estimates over the configuration space to approximate the Pareto curve.

In particular, we select only the configurations expected to be Pareto-optimal by considering as cost the weighted sum of the utilized resources. We assessed the quality of the DSEs measuring the Average Distance from Reference Set (ADRS)~\citep{schafer2020survey, ferretti2020leveraging, jhye2020transfer} metric among the real Pareto-solutions and the one estimated to be Pareto-optimal. A low value of ADRS implies a close approximation of the real Pareto-frontier.
To increase robustness to prediction errors we iteratively select candidate Pareto-optimal points by removing from the configuration space the already selected configurations and recomputing the Pareto curve up to $5$ times. 
Results are shown in Figure~\ref{fig:approaches_comparison}. In particular, we compare the performance of the fine-tuning approach against the current state-of-the-art one on the considered dataset, namely the \textit{prior-knowl.} approach~\citep{ferretti2020leveraging} (see Section~\ref{sec:related_work}). For the fine-tuning procedure we use a maximum of $128$ points from the target domain, capped at $5\%$ of the configuration space dimension; we fix the number of SGD updates to $150$, with a batch size of $32$.
We also compare the performance in the zero-shot setting~(no fine-tuning).

\begin{figure}[t]
\centering
\begin{minipage}{.48\textwidth}
\vspace{3mm}
    \centering
    \includegraphics[width=\textwidth]{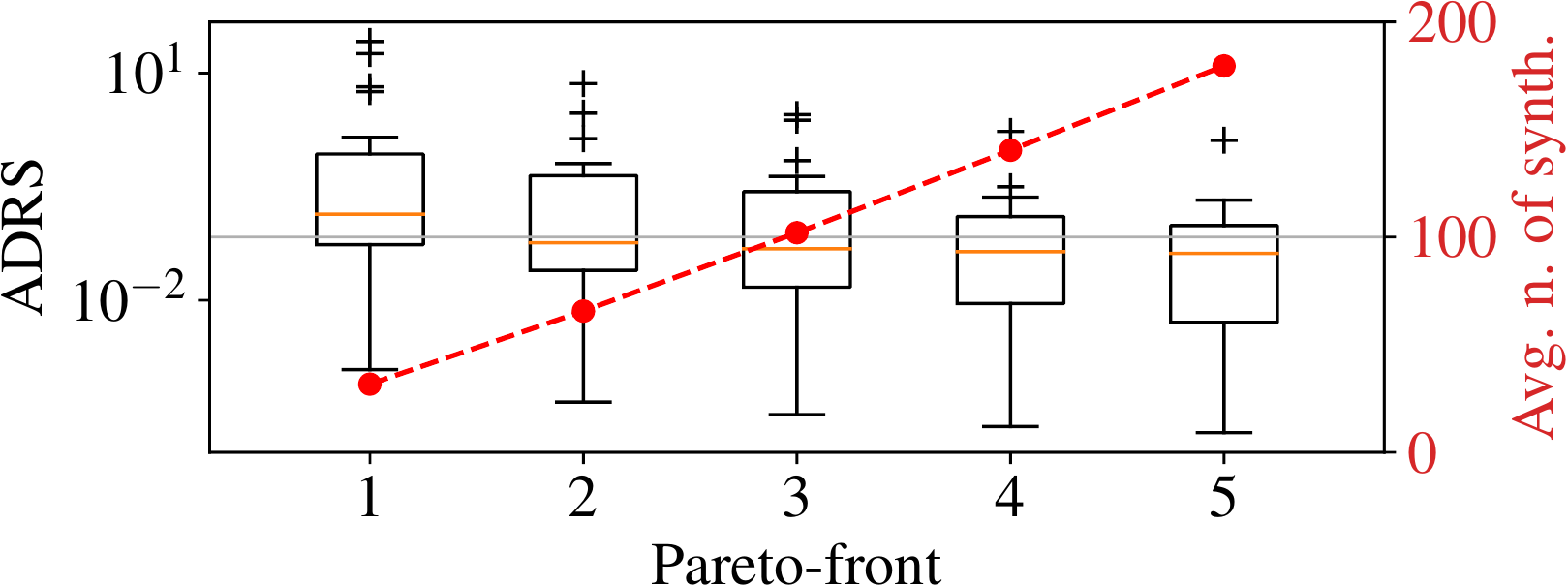}
     \vskip -0.5em
    \caption{Effect of the inference from multiple Pareto-frontier on the ADRS and the number of synthesis.}
    \label{fig:pareto_fronts}
    \vskip -1.2em
\end{minipage}%
\hspace{2mm}
\begin{minipage}{.43\textwidth}
 \centering
    \includegraphics[width=\textwidth]{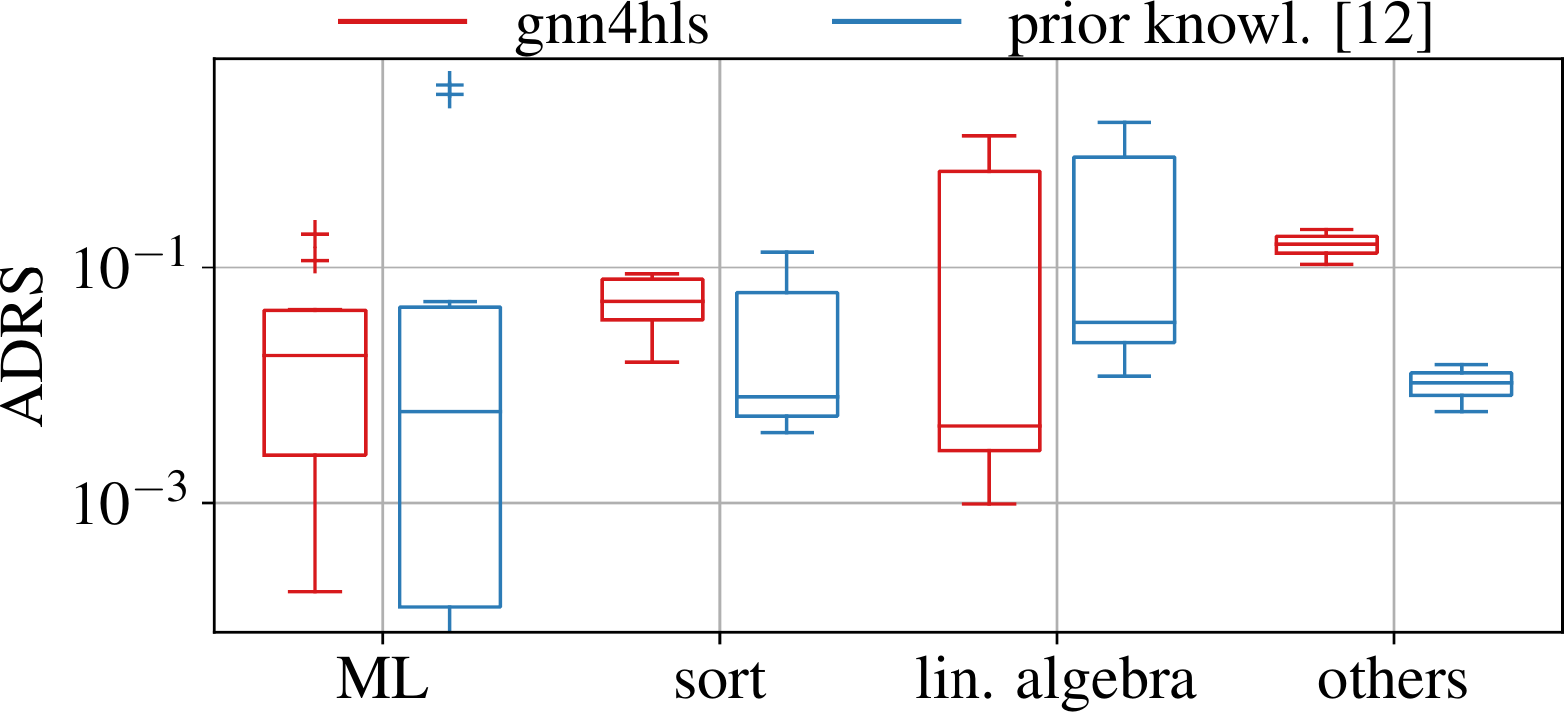}
    \vskip -0.5em
    \caption{ADRS Comparison across different class of applications among \texttt{gnn4hls} and \textit{prior knowl.}.}
    \label{fig:app_comparison}
    \vskip -1em
\end{minipage}
\end{figure}

Figure~\ref{fig:approaches_comparison} shows the performance obtained by the \textit{fine-tuned} approach~(averaged over $40$ independent runs) \wrt\ \textit{prior-knowl.} and the \textit{zero-shot} model. As previously mentioned, we consider up to the $5^{th}$ Pareto-front and compare against the reported results for \textit{prior knowl.} which iterates up to the $10^{th}$ frontier. We considered only the portion of the configuration space that is actually synthesizable~(\ie\ we considered only configurations present in the database).
Results show the distributions of the ADRS across the $23$ considered functions. In particular, the \textit{fine-tuned} model obtains Pareto-frontier approximations comparable to the state of the art, while reducing the number of outliers compared to \textit{prior knowl.} and obtaining an average ADRS of $0.20$ vs $0.45$: a remarkable improvement. 
This result is particularly appealing when observing that our approach neither uses any heuristic to perform the initial sampling, nor does it rely on domain knowledge provided by the designer. 
Furthermore, our method provides the user with performance and cost estimates of the candidate configurations, which could be instrumental to further reduce the number of syntheses required to obtain the desired performance and satisfy hardware constraints.
Figure \ref{fig:pareto_fronts} shows how the ADRS scores change \wrt\ the number of iteratively estimated Pareto frontiers during the DSE. While the ADRS decreases exponentially (plot in logarithmic scale along the $y$ axis), the number of required syntheses grows linearly. Compared to \textit{prior knowl.} our approach requires a higher number of syntheses. However, we argue that we might expect this gap to reduce significantly when considering a larger dataset: the performance of the zero-shot approach, in fact, should be considered in light of the fact that our dataset contains only $23$ different designs. 
Finally, Figure \ref{fig:app_comparison} compares the results of \texttt{gnn4hls} and \textit{prior knowl.} across the different class of applications considered in this work (see appendix for details on the taxonomy and the complete set of results). Our method shows lower variance in the ML designs, but a higher mean, and lower ADRS in linear algebra designs which are relevant for deep learning.

\section{Related works}\label{sec:related_work}

\paragraph{Design Space Exploration approaches in High-Level Synthesis}
In past years, the hardware design community has proposed different works to address the HLS-driven DSE problem. A recent survey from \citet{schafer2020survey} summarize them. Among these works we can identify two main categories: \emph{model-based} approaches, and \emph{refinement-based} ones.
Model-based approaches ~\cite{ZhongDec14, PhamMar15, ZhongJun16, ZhaoOct17, choi2017hlscope+} rely on estimates of performance and resource requirements of a given optimization. These approaches require very few synthesis runs to approximate the Pareto frontier, but, often, have difficulties while dealing with multiple, interdependent optimizations. 
Conversely, refinement-based methodologies are agnostic to the number and types of directives considered, and rely on the outcome of few heuristically sampled synthesis runs as a starting point for DSE. After the initial synthesis, the models aim to improve the initial solutions using different strategies such as genetic algorithms~\citep{SchaferMay12}, simulated annealing~\citep{mahapatra2014machine}, clustering~\citep{FerrettiJan18} or local search techniques~\citep{ferretti2018lattice}. Refinement-based approaches are not limited by the number and type of synthesis directives pre-characterized, but usually converge more slowly to the Pareto-frontier with respect to model-based approaches.
Taking a different stance, more recently \citet{ferretti2020leveraging} have proposed a DSE strategy able to map the result of past DSE targeting different design to unseen ones. The proposed approach searches for similarity in the source code of the already explored designs, and, based on the result of past DSEs, decides how to optimize the target one. This approach has been used as a baseline to compare the DSEs performance of the experiment in Section~\ref{sec:experimental_evaluation_dse}. Similarly, a recent work from \citep{jhye2020transfer} proposes a neural network model for mixed-sharing multi-domain transfer learning to transfer the knowledge obtained from previously explored design spaces in exploring a new target one.

\paragraph{Graph neural networks for hardware/software design.}First GNNs date back to models developed by~\citet{gori2005new} and~\citet{scarselli2008graph} and earlier ideas on how to process graph-structured data with recurrent neural networks~\cite{sperduti1997supervised, frasconi1998general}. In recent years, the field of graph deep learning has surged in popularity and several architectural improvements and variants of GNNs have found wide spread and adoption by the community~\cite{li2016gated, kipf2017semi,monti2017geometric, hamilton2017inductive, velickovic2018graph, bianchi2021graph}. Among their many applications to structured data processing, GNNs have been widely used as the learning system of choice in software engineering to automate program analysis and code optimization~\cite{si2018learning, li2019graph, bieber2020learning, zhou2020transferable}. Furthermore, graphs have also been used to capture the structure of hardware architectures. Notably,~\citet{mirhoseini2020chip} used a GNN to learn a transferable representation to tackle the critical hardware design problem of chip placement. Finally, due to the flexibility of graph representations, GNNs have be used to automate design processes in many other areas of science and engineering: prime examples are in the discovery of new molecules~\citep{li2018learning, you2018graph} and in automatic robot design~\citep{wang2019neural, zhao2020robogrammar}.

\section{Conclusion and future works}\label{s:conclusion}

In this work we presented \texttt{gnn4hls}, a graph-based learning framework for HLS-driven DSE. Compared against the state of the art, our method offers tools that are general and that can easily be applied to any configuration space. A key aspect of \texttt{gnn4hls} is its simplicity \wrt the its effectiveness. We show that our method compares favorably against the state of the art \wrt\ both quantitative and qualitative metrics. In the future, we plan to investigate more advanced solutions from the few-shot learning literature to improve transfer to unseen domains and to consider the a finer-grained representation of basic blocks. Then, we argue for the possibility of replacing existing heuristics to perform DSE with an exploration policy learned by~(model-based) reinforcement learning. To support breakthroughs in this direction, we renew our invitation to the community in participating in a common effort for the collection of datasets fully enabling the application of deep learning in the context of HLS-driven DSE.
We believe that this work represents a milestone for the application of graph deep learning in EDA and a significant step towards fully automated design of hardware accelerators with no human in the loop.

%% file: appendix.tex
\section{Dataset and experimental set-up}

In this section we provide details on the dataset and the experimental setting used for the experiments presented in the paper. All experiments were carried out on a server with Intel Xeon Silver CPUs and Nvidia Titan Xp/V GPUs. 

For developing the models and the infrastructure to run experiments we relied on the following open-source tools and libraries:
\begin{itemize}
    \item LLVM ~\citep{lattner2004llvm};
    \item FramaC~\citep{Couq2021framac};
    \item numpy~\citep{harris2020array};
    \item db4hls~\citep{ferretti2021db4hls};
    \item PyTorch~\citep{paszke2019pytorch};
    \item PyTorch Geometric~\citep{fey2019fast};
    \item PyTorch Lightning~\citep{falcon2019pytorch};
\end{itemize}

The code used to run the experiments is attached as supplementary material and it will be released on GitHub upon publication.

\subsection{Dataset description}\label{a:data}
The synthesis data used to train and test the model come from the $\texttt{db4hls}$ database~\citep{ferretti2021db4hls}. The database includes a collection of DSEs performed for HLS designs of the MachSuite benchmark suite~\citep{reagen2014machsuite}. In this work we use the data collected in the database according to the configuration space definition described by~\citet{ferretti2021db4hls}. The performed DSEs include up to $5$ different type of pragmas: resource type, array partitioning type, array partitioning factor, loop unrolling, and function inlining. The pragma values specified for array partitioning factor and loop unrolling  are limited to power of twos or integer divisors of the input/output array sizes and loop trip-counts. We consider only a subset of the functions available in \texttt{db4hls}. In particular, we select only HLS designs which do not require allocation of arrays or structs. This limitation is due to the LLVM compiler analysis performed to extract the CFG representation of the software specifications: this limitation will be addressed in future works. The original designs from MachSuite are compiled with the lowest optimisation level (\texttt{-O0}). Then, we process the intermediate representation (IR) produced by LLVM through a custom LLVM pass in order to generate the CFG and distinguish among the different types of block. Due to some LLVM optimisation performed on the IR representation, memory allocation are in some cases transformed into function calls generating an undesired code transformation which affect the the resulting CFG -- \ie\ additional call blocks are generated without an explicit function invocation in the original specifications. We aim at addressing this limitation in the next future.

In addition to the block type, the LLVM pass extracts block specific information. This information is used to populate the block attribute vectors used to represent each node of the hybrid control data flow graph representation. In order to extract such information, we perform different analyses of the LLVM IR, the Clang abstract syntax tree (AST) and the source code. An LLVM pass generates the intermediate representation, which is used to first identify basic blocks in the code and the traditional CFG representation. Then, for each instruction in the basic blocks we check the type of operation performed and we discriminate among the different types of block. Basic blocks including store instructions are split separating the store instruction from the predecessor and successor instructions. The newly created store block is then is connected to the new basic blocks resulting from the split. Read blocks are generated in the same way.
Loop blocks are identified from the IR. Loop information is extracted in order to model loop carried dependencies, loop stride, and loop trip count.
Similarly, call blocks are identified from the IR. In this case information about the number of parameters, number of function invocation, and number of LLVM instructions of the invoked function are extracted and included in the attribute vector.
From the remaining uncategorized standard blocks, information about the number of LLVM instructions executed in it are extracted and added to the attribute vector.

Parameter blocks instead are generated combining information from the LLVM IR and the AST analysis. In particular we extract information associated to the parameter type (pointer or value), parameter data type -- the size of the data type --, and number of elements of pointer parameters. This last information is required by the HLS tools in order to know the memory required by the hardware accelerator. However, since this information is not preserved in the LLVM IR, we have extracted it from a joint AST and source code analysis. To avoid large numerical differences among attributes, we use logarithmic scale fot the number of LLVM instructions, loop trip counts, array partitioning factors, data types, and their related directive values.

Table~\ref{tab:list_of_functions} reports the designs considered in this work. 
For each function we show: the application domain (used to group the applications in Figure~\ref{fig:app_comparison}), the original benchmark name in MachSuite, the HLS design name, the size of the designs in term of lines of code, the type of pragmas considered during the exploration and the configuration space size.

\begin{table}[]
\caption{List of the functions considered in this work. The table shows: application domain, the original benchmark from MachSuite \citep{reagen2014machsuite}, the target HLS design to implement in hardware, number of lines of code, type of pragmas considered for the exploration, configuration space size. Pragma types are indicated by different symbols: resource $\blacktriangle$, array partition type $\blacktriangledown$, array partition factor $\blacksquare$, loop unrolling $\blacklozenge$, function inlining $\bigstar$.\label{tab:list_of_functions}}
\centering
\resizebox{\textwidth}{!}{
\begin{tabular}{@{}llllll@{}}
\toprule
Domain             & Benchmark  & HLS design                                         & Lines of code & Type of pragmas                                                                                           & | CS | \\ \midrule
\rowcolor{Gray}
&       & ncubed                                             & 41            & $ \blacktriangledown \quad \blacksquare \quad \blacklozenge \quad$                                                                                                  & 2744   \\
\rowcolor{Gray}
                   &     \multirow{-2}{*}{gemm}       & bbgemm                                             & 45            &         $ \blacksquare \quad \blacklozenge \quad$                                                                                                  & 1600   \\
\rowcolor{Gray}
\multirow{-3}{*}{\shortstack[l]{Linear\\Algebra}}                        & spmv       & ellpack                                            & 28            &                                                                                     $ \blacksquare \quad \blacklozenge $                      & 1600    \\
\multirow{7}{*}{Sorting}            & \multirow{7}{*}{radix sort} & hist                                               & 34            &      $\blacktriangledown \quad \blacksquare \quad \blacklozenge$                                                                                                      & 4704     \\
                   &            & init                                               & 29            &         $\blacktriangle \quad \blacktriangledown \quad \blacksquare \quad \blacklozenge \quad$                                                                                                  & 484   \\
                   &            & last\_step\_scan                                   & 32            &                               $\blacktriangledown \quad \blacksquare \quad \blacklozenge$                                                                            & 800  \\
                   &            & sum\_scan                                          & 31            &                               $\blacktriangle \quad \blacktriangledown \quad \blacksquare \quad \blacklozenge$                                                                            & 1280      \\
                   &            & local\_scan                                        & 32            &                               $\blacktriangledown \quad \blacksquare \quad \blacklozenge$                                                                            & 704   \\
                   &            & update                                             & 36            &                               $\blacktriangledown \quad \blacksquare \quad \blacklozenge$                                                                            & 2400    \\
                   &            & ss\_sort                                           & 50            &                               $\blacksquare \quad \blacklozenge \quad \bigstar $                                                                            & 1792   \\
\rowcolor{Gray}                   
\multirow{2}{*}{Stencil}            &    & stencil2d                                          & 39            &   $\blacktriangledown \quad \blacksquare \quad \blacklozenge \quad$ & 1344      \\
\rowcolor{Gray}
 \multirow{-2}{*}{Stencil}                   &   \multirow{-2}{*}{stencil}          & stencil3d                                          & 62            &  $ \blacktriangledown \quad \blacksquare \quad \blacklozenge \quad$                                                                                                         & 1536      \\
Biology & md         & knn                                                & 71            &              $\blacksquare \quad \blacklozenge$                                                                                             & 1152   \\

\rowcolor{Gray}
  &    & get\_delta\_matrix\_weights1                       & 45            &     $\blacktriangledown \quad \blacksquare \quad \blacklozenge$                                                                                                      & 21952   \\
  \rowcolor{Gray}
                   &            & get\_delta\_matrix\_weights2                       & 45            &        $\blacktriangledown \quad \blacksquare \quad \blacklozenge$                                                                                                   & 31213  \\
\rowcolor{Gray}                   
                   &            & get\_delta\_matrix\_weights3                       & 45            &         $\blacktriangledown \quad \blacksquare \quad \blacklozenge$                                                                                                  & 21952  \\
\rowcolor{Gray}                   
                   &            & product\_with\_bias\_input\_layer  & 48            &         $\blacksquare \quad \blacklozenge \quad \bigstar$                                                                                                  & 1372 \\
\rowcolor{Gray}                   
                   &            & product\_with\_bias\_second\_layer & 48            &                                               $\blacksquare \quad \blacklozenge \quad \bigstar$                                                             & 686  \\
\rowcolor{Gray}                   
                   &            & product\_with\_bias\_output\_layer & 48            &          $\blacksquare \quad \blacklozenge \quad \bigstar$                                                                                                  & 392 \\
\rowcolor{Gray}                   
                   &            & take\_difference                                   & 44            &                                $\blacktriangledown \quad \blacksquare \quad \blacklozenge$                                                                            & 512\\
\rowcolor{Gray}                   
                   &            & get\_oracle\_activations1                          & 48            &                                  $\blacksquare \quad \blacklozenge$                                                                         & 2401   \\
\rowcolor{Gray}                   
                   &            & get\_oracle\_activations2                          & 48            &                                 $\blacksquare \quad \blacklozenge$                                                                          & 1372\\
\rowcolor{Gray}                   
 \multirow{-10}{*}{\shortstack[l]{Machine\\Learning}}                   &     \multirow{-10}{*}{backprop}     & update\_weights                                    & 127           &        $\blacksquare \quad \blacklozenge$                                                                                                   & 1024 \\ \bottomrule
\end{tabular}
}
\end{table}

\subsection{Global attributes}

In addition to the attribute vectors associated to each node, we introduce a global representation of the configuration with respect to its configuration space. 

Given a design and its associated configuration space, we define $\vs \in \sR^5$ as the vector having for each component the average among the directive values set associated to each pragma type.
As an example, given a configuration space having directive value sets with 2 resource types~(one-hot encoded), 2 partitioning types~(one-hot encoded), partitioning factors of $1$, $2$, $4$, $8$, unrolling factor of $10$, $20$, $30$, and 2 options for function inlining~(one-hot encoded), the resulting configuration space vector will be $\vs$ = $[0.5, 0.5, 3.75, 20, 0.5]$.
Then, given a specific configuration, we generate the vector $\vc \in \sR^5$ as the average vector among directive values of the same type in that particular configuration. Similarly, we generate vectors $\vs^\prime$ and $\vc^\prime$ using the median among the directive set values instead of the mean. 
Finally, given the number of instructions in the function $l$ and the number of input parameters $p$ we define the global attribute vector of the configuration $\vu$ as:
\begin{equation}
    \vu = l \concat p \concat (\vs - \vc) \concat (\vs^\prime - \vc^\prime),
\end{equation}
where ${}\concat{}$ is the vector concatenation operation. 
Intuitively, this vector representation captures how much a configuration leans toward a region of the design space. In practice, the vectors $(\vs - \vc)$ and $(\vs^\prime - \vc^\prime)$ are also normalized so that each element has unitary variance across the configuration space.

\subsubsection{Design space exploration performance metric}
The results of the DSEs have been evaluated in term of Average Distance from Reference Set (ADRS). The ADRS metric is used to quantify the distance between a reference curve $P$, and an approximated one $\bar{P}$. In our case, the reference curve is the Pareto-frontier ground truth available computed over the synthesized design avaiable in the dataset, while the approximated one is the Pareto-frontier resulting from the DSE performed by using the trained model.
The ADRS for two objective functions is defined as:
\begin{align}
	&ADRS(\bar{P}, P) = \bigg[ \frac{1}{|P|} \sum\limits_{p \in P} \min_{\bar{p}\in \bar{P}} (d(\bar{p}, p)) \bigg], \\
	&d(\bar{p}, p) = \max \{0,(A_{\bar p} - A_{p})/A_{p},(L_{\bar p} - L_{p})/L_{p}\}
\end{align}
where $A_{\bar p}$ and $L_{\bar p}$ are the area and latency of an element of the reference Pareto-frontier, while $A_{p}$ and $L_{p}$ are the area and latency of the approximated one.
Intuitively, lower ADRS values implies proximity among the approximated curve and the reference one. We consider as altency~($\mathit{LAT}$) the number of clock cycles required by the hardware implementation to execute the functionality, and, as measure of the area ($A$), the number aggregated values of $\mathit{FF}$, $\mathit{LUT}$, $\mathit{DSP}$ in form of a linear combination of their utilisation.

\begin{equation}
	A = \frac{\mathit{FF}}{\mathit{FF}_{available}} + \frac{\mathit{LUT}}{\mathit{LUT}_{available}} + \frac{\mathit{DSP}}{\mathit{DSP}_{available}} 
\end{equation}\label{eq:aggregated_area}

This formulation allows to obtain a unique metric for the area costs evaluating the overall utilisation of the resources required by an implementation ($\mathit{FF}$, $\mathit{LUT}$, and $\mathit{DSP}$) with respect to the ones available on a specific FPGA ($\mathit{FF}_{available}$, $\mathit{LUT}_{available}$, and $\mathit{DSP}_{available}$).

\section{Hyperparameters and additional results}

In this section first we provide details on the hyperparameters and architecture used for the different models and how training was performed, then we show additional~(more complete) experimental results.

\subsection{Performance and cost estimation experiment}

We use a GNN with $4$ propagation blocks. All the MLPs required to implement encoding, update and message functions are implemented as networks with a single hidden layer and $\texttt{ELU}$ activation function~\citep{clvert2016elu}. The MLPs processing the node representations all have $128$ hidden units, while we use a width of $256$ to process global graph features, both in the propagation blocks and in the regression head. The MLPs computing node messages ($\texttt{MLP}^t_{\psi_\vv}$) have and additional linear layer after the nonlinear one. Finally, the MLP used to compute attention scores has an hidden layer of $256$ units and a number of output units equal to the number of attention heads with $\texttt{LeakyReLU}$~($0.2$ negative slope) activation as in~\citep{velickovic2018graph}. We use $2$ attention heads in parallel and we concatenate their outputs before processing with the global update function $\tau_\vu^t$. For the baseline we use a node-level MLP with $5$ hidden layers with $512$ units each and $\texttt{ReLU}$ activation function, followed by $\texttt{SUM}\{{}\cdot{}\}$ aggregation and a second global MLP with a single hidden layer with the same activation and  number of neurons.
 
For training, we use as target values the natural logarithm of the true values, and the mean absolute error as loss function. Models are trained for $800$ epochs with a batch-size of $128$ without early stopping and by taking as final model the one that achieves the lowest validation error across the training epochs. For optimization we use the Adam optimizer~\cite{kingma2015adam} with an initial learning rate of $0.001$ and cosine annealing (with no restarts) as a schedule with a minimum learning rate of $0.0001$. We also clip the gradient norm to a maximum value of $3$ to avoid learning instabilities. We did not use any form of regularization since we did not observe any sign of overfitting. We do not perform any additional scaling to the input features w.r.t. the preprocessing steps described in~\ref{a:data}.
 
Table~\ref{tab:regression_gnnhls} and Table~\ref{tab:regression_baseline} show the estimation accuracy for \texttt{gnn4hls} and the DeepSets \texttt{baseline} respectively. In particular, we show performance in terms of Mean Absolute Error~(MAE) and Mean Absolute Percentage Error~(MAPE) for all the prediction targets for each design. Note that the MAPE of the DSP estimate is not defined for some function where DSP is not used, tables report a $0.0\%$ error for both models in such cases. Results are averaged over $5$ independent runs, we do not report standard deviations for the sake of presentation clarity.
 
 \begin{table}[ht]
    \caption{Detailed results of the performance and costs estimation experiment for $\texttt{gnnhls}$.\label{tab:regression_gnnhls}}
    \resizebox{\textwidth}{!}{\begin{tabular}{@{}lllllllll@{}}
    \toprule
    & \multicolumn{8}{c}{\texttt{gnnhls}}\\ 
    & \multicolumn{4}{c}{MAPE}& \multicolumn{4}{c}{MAE}\\ \cmidrule(l){2-5}\cmidrule(l){6-9}
\multirow{-1}{*}{HLS design}&\multicolumn{1}{c}{LAT}&\multicolumn{1}{c}{FF}&\multicolumn{1}{c}{LUT}&\multicolumn{1}{c}{DSP}&\multicolumn{1}{c}{LAT}&\multicolumn{1}{c}{FF}&\multicolumn{1}{c}{LUT}&\multicolumn{1}{c}{DSP}\\ \midrule
ncubed	&1.9\%	&3.67\%	&3.48\%	&3.18\%	&2.37E+05	&6.06E+02	&1.21E+03	&4.80E+00	\\ 
bbgemm	&1.35\%	&2.8\%	&2.24\%	&0.46\%	&2.42E+05	&3.13E+02	&2.58E+02	&1.35E-01	\\ 
ellpack	&2.69\%	&2.58\%	&2.41\%	&2.21\%	&1.14E+04	&8.04E+02	&6.08E+02	&6.37E+00	\\ 
hist	&3.68\%	&8.54\%	&5.41\%	&0.0\%	&6.53E+02	&6.82E+02	&1.64E+03	&1.73E-03	\\ 
init	&2.86\%	&15.03\%	&4.7\%	&0.0\%	&4.19E+01	&4.04E+01	&3.78E+02	&1.95E-03	\\ 
last\_step\_scan	&8.31\%	&35.95\%	&8.0\%	&0.0\%	&6.77E+02	&7.72E+02	&2.47E+03	&2.02E-03	\\ 
sum\_scan	&0.8\%	&1.19\%	&1.05\%	&0.0\%	&5.31E+00	&1.05E+01	&3.53E+01	&1.22E-03	\\ 
local\_scan	&6.5\%	&11.61\%	&7.15\%	&0.0\%	&8.94E+02	&7.87E+02	&1.92E+03	&1.23E-03	\\ 
update	&1.41\%	&4.49\%	&2.35\%	&0.0\%	&4.45E+02	&1.84E+02	&6.11E+02	&8.17E-04	\\ 
ss\_sort	&0.68\%	&1.24\%	&1.27\%	&0.0\%	&1.37E+04	&5.42E+01	&3.47E+02	&1.40E-03	\\ 
stencil2d	&3.75\%	&3.62\%	&2.57\%	&6.94\%	&1.19E+04	&3.07E+02	&9.08E+02	&3.01E+01	\\ 
stencil3d	&4.21\%	&8.65\%	&6.35\%	&6.68\%	&1.74E+04	&1.46E+03	&2.75E+03	&1.46E+01	\\ 
knn	&0.99\%	&1.6\%	&1.72\%	&1.38\%	&8.17E+03	&3.12E+02	&6.50E+02	&1.68E+00	\\ 
get\_delta\_matrix\_weights1	&0.62\%	&0.83\%	&0.69\%	&0.61\%	&1.97E+02	&8.47E+01	&1.12E+02	&2.04E+00	\\ 
get\_delta\_matrix\_weights2	&1.51\%	&2.45\%	&2.04\%	&1.36\%	&1.40E+03	&4.48E+02	&5.05E+02	&1.35E+01	\\ 
get\_delta\_matrix\_weights3	&0.59\%	&0.78\%	&0.65\%	&0.4\%	&7.35E+01	&2.92E+01	&3.20E+01	&7.06E-01	\\ 
product\_with\_bias\_input\_layer	&0.67\%	&0.82\%	&1.5\%	&0.56\%	&3.61E+02	&6.62E+01	&1.61E+02	&7.54E-01	\\ 
product\_with\_bias\_second\_layer	&0.31\%	&0.28\%	&0.27\%	&0.18\%	&1.22E+03	&7.71E+00	&3.38E+01	&2.82E-02	\\ 
product\_with\_bias\_output\_layer	&0.29\%	&0.29\%	&0.38\%	&0.2\%	&5.23E+01	&3.94E+00	&7.29E+00	&3.18E-02	\\ 
take\_difference	&0.35\%	&0.18\%	&0.13\%	&0.21\%	&6.10E-01	&2.53E+00	&2.27E+00	&4.91E-02	\\ 
get\_oracle\_activations1	&2.48\%	&2.7\%	&2.79\%	&2.67\%	&4.32E+03	&1.70E+02	&3.19E+02	&1.74E+00	\\ 
get\_oracle\_activations2	&2.42\%	&1.46\%	&2.13\%	&1.85\%	&1.93E+02	&8.76E+01	&1.51E+02	&1.46E+00	\\ 
update\_weights	&0.39\%	&0.72\%	&0.83\%	&0.25\%	&7.76E+03	&5.13E+01	&1.05E+02	&3.48E-02	\\ 
\midrule
\textbf{Avg.}	&\textbf{2.12\%}	&\textbf{4.85\%}	&\textbf{2.61\%}	&\textbf{1.27\%}	&\textbf{2.43E+04}	&\textbf{3.17E+02}	&\textbf{6.61E+02}	&\textbf{3.39E+00}	\\ \bottomrule
\end{tabular}}
    \end{table}

\begin{table}[ht]
    \caption{Detailed results of the performance and costs estimation experiment for the $\texttt{baseline}$.\label{tab:regression_baseline}}
    \resizebox{\textwidth}{!}{\begin{tabular}{@{}lllllllll@{}}
    \toprule
    & \multicolumn{8}{c}{\texttt{baseline}}\\ 
    & \multicolumn{4}{c}{MAPE}& \multicolumn{4}{c}{MAE}\\ \cmidrule(l){2-5}\cmidrule(l){6-9}
\multirow{-1}{*}{HLS design}&\multicolumn{1}{c}{LAT}&\multicolumn{1}{c}{FF}&\multicolumn{1}{c}{LUT}&\multicolumn{1}{c}{DSP}&\multicolumn{1}{c}{LAT}&\multicolumn{1}{c}{FF}&\multicolumn{1}{c}{LUT}&\multicolumn{1}{c}{DSP}\\ \midrule
ncubed	&8.24\%	&29.81\%	&17.44\%	&29.44\%	&7.72E+05	&4.39E+03	&4.57E+03	&2.12E+01	\\ 
bbgemm	&5.34\%	&11.26\%	&7.05\%	&4.39\%	&9.32E+05	&1.32E+03	&8.01E+02	&1.20E+00	\\ 
ellpack	&29.99\%	&36.96\%	&21.09\%	&92.78\%	&1.50E+05	&5.86E+03	&3.67E+03	&5.13E+01	\\ 
hist	&18.67\%	&49.28\%	&43.0\%	&0.0\%	&2.64E+03	&1.79E+03	&8.35E+03	&2.96E-02	\\ 
init	&8.47\%	&53.91\%	&273.64\%	&0.0\%	&9.41E+01	&1.94E+02	&6.10E+02	&5.36E-02	\\ 
last\_step\_scan	&27.21\%	&47.21\%	&12.9\%	&0.0\%	&1.64E+03	&1.52E+03	&3.15E+03	&3.97E-02	\\ 
sum\_scan	&4.32\%	&7.61\%	&3.61\%	&0.0\%	&3.08E+01	&5.21E+01	&1.08E+02	&1.84E-02	\\ 
local\_scan	&17.13\%	&14.32\%	&8.67\%	&0.0\%	&1.93E+03	&1.15E+03	&2.23E+03	&4.92E-02	\\ 
update	&6.46\%	&19.73\%	&12.77\%	&0.0\%	&2.05E+03	&6.72E+02	&2.78E+03	&2.02E-02	\\ 
ss\_sort	&1.79\%	&2.47\%	&2.43\%	&0.0\%	&3.56E+04	&1.14E+02	&6.63E+02	&8.55E-03	\\ 
stencil2d	&12.35\%	&7.82\%	&5.51\%	&19.29\%	&3.56E+04	&7.70E+02	&1.92E+03	&6.65E+01	\\ 
stencil3d	&23.16\%	&25.92\%	&20.36\%	&41.69\%	&6.96E+04	&4.00E+03	&8.97E+03	&6.51E+01	\\ 
knn	&3.02\%	&4.92\%	&20.33\%	&2.8\%	&1.80E+04	&8.60E+02	&8.35E+03	&3.44E+00	\\ 
get\_delta\_matrix\_weights1	&5.1\%	&8.23\%	&6.67\%	&9.1\%	&1.49E+03	&6.27E+02	&7.78E+02	&1.58E+01	\\ 
get\_delta\_matrix\_weights2	&40.05\%	&31.02\%	&32.77\%	&60.15\%	&2.11E+04	&2.93E+03	&8.26E+03	&8.60E+01	\\ 
get\_delta\_matrix\_weights3	&3.59\%	&6.68\%	&4.55\%	&4.12\%	&4.25E+02	&2.67E+02	&2.39E+02	&6.76E+00	\\ 
product\_with\_bias\_input\_layer	&1.53\%	&2.23\%	&3.91\%	&1.39\%	&8.43E+02	&1.74E+02	&4.84E+02	&1.20E+00	\\ 
product\_with\_bias\_second\_layer	&1.24\%	&1.69\%	&5.28\%	&1.36\%	&4.99E+03	&3.43E+01	&7.76E+02	&2.11E-01	\\ 
product\_with\_bias\_output\_layer	&0.79\%	&4.08\%	&3.89\%	&0.5\%	&1.44E+02	&6.12E+01	&8.85E+01	&7.73E-02	\\ 
take\_difference	&2.26\%	&1.22\%	&1.17\%	&1.29\%	&3.48E+00	&2.60E+01	&3.07E+01	&4.72E-01	\\ 
get\_oracle\_activations1	&4.74\%	&5.76\%	&5.75\%	&5.73\%	&9.97E+03	&5.66E+02	&1.09E+03	&4.23E+00	\\ 
get\_oracle\_activations2	&8.85\%	&4.46\%	&5.96\%	&4.37\%	&6.97E+02	&3.30E+02	&4.60E+02	&4.39E+00	\\ 
update\_weights	&1.18\%	&1.5\%	&5.17\%	&0.31\%	&2.33E+04	&1.07E+02	&6.40E+02	&4.38E-02	\\ 
\midrule
\textbf{Avg.}	&\textbf{10.24\%}	&\textbf{16.44\%}	&\textbf{22.78\%}	&\textbf{12.12\%}	&\textbf{9.06E+04}	&\textbf{1.21E+03}	&\textbf{2.56E+03}	&\textbf{1.43E+01}	\\ \bottomrule
\end{tabular}}
    \end{table}

\begin{table}[]
\caption{DSE result comparison among \texttt{gnn4hls} and \textit{prior-knowl}.\label{tab:datailed_dse_results}}
\resizebox{\textwidth}{!}{\begin{tabular}{@{}lllll@{}}
\toprule
      \multirow{4}{*}{HLS design}                                             & \multicolumn{2}{c}{\texttt{gnn4hls}} & \multicolumn{2}{c}{\textit{prior-knowl}}       \\ \cmidrule(l){2-5} 
                                            & ADRS     & \# of syntehsis  & ADRS                         & \# of synthesis \\ \midrule
ncubed                                             & 0.043 $\pm$ 0.042    & 146  $\pm$ 27             & 0.012                        & 35              \\
bbgemm                                             & 0.016 $\pm$ 0.018     & 148 $\pm$ 29          & 0.007                        & 46               \\
ellpack                                            & 0.032 $\pm$ 0.02    & 115   $\pm$ 21           & 0.034                        & 65              \\
hist                                               & 0.107 $\pm$ 0.199     & 119 $\pm$ 20          & 0.007                        & 46              \\
init                                               & 1.3 $\pm$ 2.36   & 61 $\pm$ 30             & 0.078                        & 68              \\
last\_step\_scan                                   & 0.211 $\pm$ 0.166   & 104    $\pm$ 19          & 0.004                        & 90              \\
sum\_scan                                          & 0.030 $\pm$ 0.02   & 143    $\pm$ 17          & 0.136                        & 25              \\
local\_scan                                        & 0.167 $\pm$ 0.24   & 88   $\pm$ 10           & 0.005                        & 71              \\
update                                             & 0.002 $\pm$ 0.002   & 97  $\pm$ 17            & 0.009                        & 28              \\
ss\_sort                                           & 0.042 $\pm$ 0.02   & 42    $\pm$ 10            & 0.0005                       & 21              \\
stencil2d                                          & 0.193  $\pm$ 0.24  & 90   $\pm$ 20           & 0.015                        & 46              \\
stencil3d                                          & 0.115 $\pm$ 0.19   & 97    $\pm$ 28           & 1.88                         & 16              \\
knn                                                & 0.006  $\pm$ 0.004  & 284  $\pm$ 42            & 0.006                        & 25              \\
get\_delta\_matrix\_weights1                       & 0.087 $\pm$ 0.036   & 427    $\pm$ 73           & 0.002                        & 139             \\
get\_delta\_matrix\_weights2                       & 0.054  $\pm$ 0.017  & 525     $\pm$ 71         & 0.010                        & 77              \\
get\_delta\_matrix\_weights3                       & 0.087  $\pm$ 0.037  & 568  $\pm$ 88            & 0.030                        & 222             \\
product\_with\_bias\_input\_layer  & 0.005  $\pm$ 0.005  & 215  $\pm$ 36             & 3.560                        & 3               \\
product\_with\_bias\_second\_layer & 0.0001  $\pm$ 0.001  & 49   $\pm$ 18            & 0                            & 30              \\
product\_with\_bias\_output\_layer & 0.003 $\pm$ 0.013   & 71   $\pm$ 32            & 2.5E-5 & 24              \\
take\_difference                                   & 0.002  $\pm$ 0.005   & 224   $\pm$ 56           & 0.0002                       & 8               \\
get\_oracle\_activations1                          & 0.048 $\pm$ 0.022   & 220   $\pm$  29         & 2.907                        & 67              \\
get\_oracle\_activations2                          & 0.013 $\pm$ 0.009   & 244    $\pm$ 41           & 0.051                        & 19              \\
update\_weights                                    & 0.0002 $\pm$ 0.0002  & 63   $\pm$ 24            & 1.1E-5 & 3               \\ \bottomrule
\end{tabular}}
\end{table}

\subsection{Design space exploration experiment}
 
For the DSE experiments we pretrained a model for each one of the available functions using a leave-one-out approach. Then we finetuned each pretrained model on the target domain as described in Section~\ref{sec:experimental_evaluation_dse}. The model architecture here is the same used for the previous experiment.  In the fine-tuning stage we used Adam with a constant learning rate of $0.001$ and clipping the gradient norm to $10$. To evaluate the impact of the initial random sampling the finetuning runs were repeated $20$ times with different random seeds.

Table \ref{tab:datailed_dse_results} reports the DSEs result obtained by the \texttt{gnn4hls} framework compared against the \textit{prior-knowl.} approach from~\citet{ferretti2020leveraging}. The table lists, for all the considered function, the ADRS values obtained and the number of synthesis required by the methodologies. For \texttt{gnn4hls} results are averaged over $40$ different runs and we report the standard deviations. 